\documentclass[11pt]{article}

\usepackage{amsmath}
\usepackage[paper=a4paper,text={16cm,24cm},centering]{geometry}
\usepackage{graphicx}

\def\be{\begin{equation}}
\def\ee{\end{equation}}

\def\gc[#1]#2{\gamma_{#1c}^{(#2)}}

\begin{document}

\title{Fluctuations of the multiplicity of produced particles
in onium-nucleus collisions}

\author{Tseh Liou$^1$, A. H. Mueller$^1$, S. Munier$^2$\\
\\
{\small\it $^1$ Department of Physics, Columbia University, New York, USA}\\
{\small\it $^2$ Centre de Physique Th\'eorique, \'Ecole Polytechnique, CNRS,}\\
{\small\it Universit\'e Paris-Saclay, Palaiseau, France.}
}

\maketitle

\begin{abstract}
We address the general features of event-by-event fluctuations of the multiplicity of 
gluons produced in the scattering of a dilute hadron off a large nucleus
at high energy in the fragmentation region of the dilute hadron.
We relate these fluctuations to the stochasticity of 
the number of quanta contained in the hadron at the time of the interaction.
For simplicity, we address the ideal case in which the hadron is an onium, and
investigate different kinematical regimes in rapidity and onium size.
We show that at large rapidity, the multiplicity distribution 
exhibits an exponential tail in the large-multiplicity
region, which is qualitatively consistent with the proton-nucleus data.
But interestingly enough, the exponential shape is determined by confinement.
\end{abstract}

\section{Introduction}

The large amount of data collected at the RHIC and at the LHC has made accessible the
study of event-by-event fluctuations of a number of measurable quantities,
such as particle multiplicities in proton-nucleus and nucleus-nucleus
collisions. 
The microscopic origin of the observed stochasticity is however not clear, and various
interpretations and phenomenological models have been proposed.\\

In the available models to date,
the stochasticity is often correlated to the 
event-by-event fluctuations of the matter density in the nucleus.
(For a recent review on quantum fluctuations in
the initial state of heavy-ion collisions, see e.g.~\cite{Gelis:2016upa}).
A common assumption is that the nucleus is a set of nucleons
whose positions in the transverse plane relative to its center 
are random, see for example the PHOBOS Glauber Monte Carlo model~\cite{Alver:2008aq}. 
The flow of particles which go to the final state is then related
to the geometry of the initial nucleus.\\

Other recent models assume that the density of gluons significantly 
fluctuates from nucleon to nucleon, and that these
fluctuations are encoded in the large-rapidity fluctuations of the saturation
scale~\cite{McLerran:2015lta}. However, such effects were shown, 
theoretically~\cite{Dumitru:2007ew} as well as phenomenologically~\cite{Gelis:2006bs},
to be small at realistic collider energies.\\

In this paper, we investigate the assumption that the fluctuations of the multiplicity of the produced particles
in the forward rapidity region of the proton in proton-nucleus collisions is entirely due to the event-by-event
fluctuations of the gluon content in the proton generated by the small-$x$ evolution, which
are known to be large (see e.g. the recent work of Ref.~\cite{PhysRevLett.117.052301}). 
The nucleus is instead a non-fluctuating object.
Indeed, we observe that generally speaking,
it is quite unnatural to have configurations 
of nucleons inside the nucleus which are very different from uniformly distributed
since the wavefunction of a large nucleus is approximately the same as the
ground-state wavefunction of nuclear matter, for which it is known that density
fluctuations are very small \cite{blaizot}.
\\

Our paper is organized as follows. In the next section, we expose our picture
of multiplicity fluctuations, relating them to the parton number fluctuations in the initial state 
of the dilute object. 
In Sec.~\ref{sec:background}, we provide the necessary background on
small-$x$ evolution in the Balitsky-Fadin-Kuraev-Lipatov (BFKL) 
\cite{Lipatov:1976zz,Kuraev:1977fs,Balitsky:1978ic} regime, and in particular, we
rederive the set of equations obeyed by the moments of the
gluon number in the framework of the color dipole model~\cite{Mueller:1993rr}.
In Sec.~\ref{sec:collinear}, we review the collinear, or double-logarithmic
(DL) limit of these equations, relevant when the rapidity is large compared to the logarithm of the
ratio of the relevant transverse scales, and solve them to obtain the multiplicity distribution.
In Sec.~\ref{sec:BFKL}, we release the DL approximation:
The result of this plain BFKL calculation will prove unphysical, 
a problem that we shall address in the subsequent section~\ref{sec:BFKLconf}.
The concluding section contains some prospects, while two appendices discuss
small-$x$ evolution in the diffusive approximation~(Appendix~\ref{app:branchingdiffusion}),
and a numerical study of a dimensionally-reduced model which
should share the main features of the full BFKL evolution~(Appendix~\ref{app:num}).


\section{\label{sec:picture}Picture of the multiplicity fluctuations}

As announced in the Introduction, 
we shall concentrate on particle production in onium-nucleus collisions
in the fragmentation region of the onium. We will argue that
in this process, the shape of the multiplicity
distribution in a specific kinematical region that we
shall define properly below can be traced back to the fluctuations
of the {\it number of quanta} in the wavefunction of the onium
at the time of its interaction with the nucleus~\cite{mueller}.
Our aim is not to build a realistic model for real hadron-nucleus
scattering, but rather to study in as many details as possible
onium (namely dipole)-nucleus scattering and derive, in this simple case, the
distribution of the number of particles (actually gluons) in the final state.
We expect that the main characteristics of the distribution we will find go
over unchanged to the experimentally measurable proton-nucleus 
or deuterium-nucleus scattering processes.
\\

This very kind of statistical fluctuations of the partonic content
of hadrons has been considered before: Their impact on the shape
and energy evolution of deep-inelastic scattering cross sections
was investigated in the deep saturation regime~\cite{Iancu:2004es}
and at more moderate energies~\cite{Mueller:2014fba} (for
reviews, see~\cite{Munier:2009pc,Munier:2014bba}).
Here, we consider the effect of these fluctuations on a final-state observable.\\

\begin{figure}
\begin{center}
\includegraphics[width=0.9\textwidth]{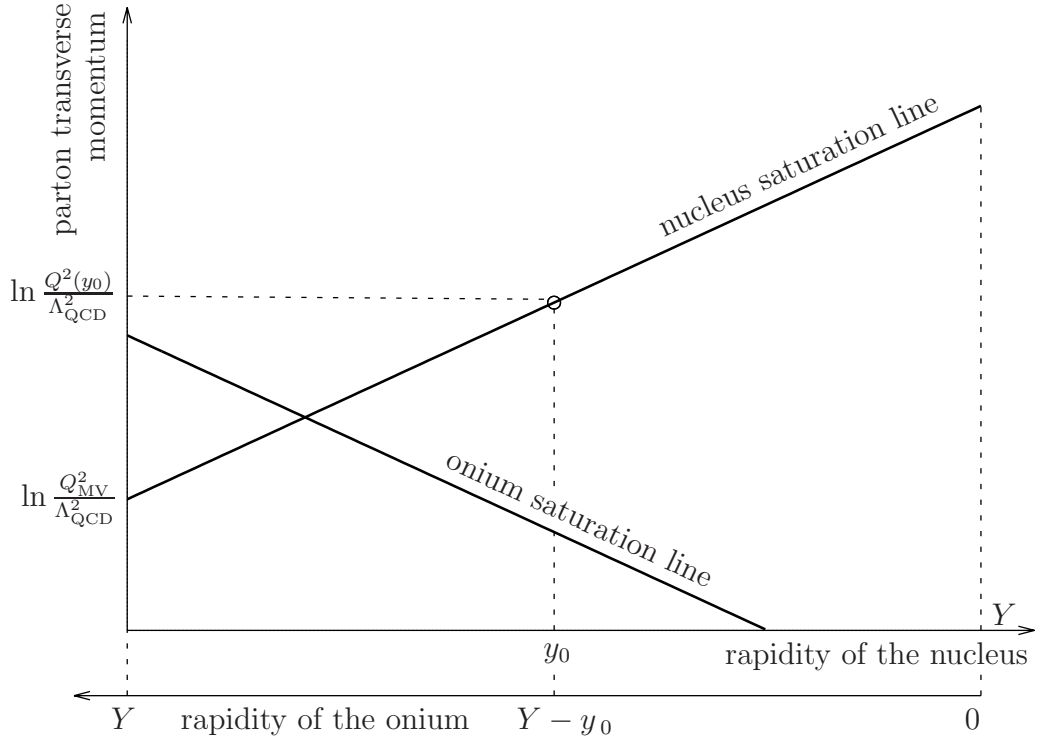}
\end{center}
\caption{\label{fig:plotintro}
\small
Kinematics of onium-nucleus scattering at fixed total
rapidity $Y$. The rapidity of the nucleus ($x$-axis) defines
the frame. The saturation lines of the onium and of the nucleus
are shown. We choose a frame, in which the nucleus has the rapidity $y_0$
along the negative $z$ axis, such that the onium is a dilute
object for gluons of transverse momenta of the order of the
saturation scale of the nucleus: The rapidity dependence
of its gluon content is then given by the small-$x$ gluon branching 
process without
saturation effects.
}
\end{figure}

Throughout our work,
the nucleus will be characterized by one single momentum scale 
$Q_s(y_0)\gg\Lambda_\text{QCD}$, its
saturation scale, which depends on its rapidity $y_0$
in the considered frame (see Fig.~\ref{fig:plotintro}).
The latter scale just sets the upper 
bound on the transverse momentum of the gluons
in the onium wavefunction that are freed and go to the final state.\\

Indeed, for an onium moving along the positive $z$ axis of a frame
in which it has the lightcone momentum $p_+$
and in the corresponding $A_+=0$ gauge, particles are produced as follows:\footnote{%
See e.g. Ref.~\cite{Kovchegov:2012mbw} for a good review on 
high-energy scattering and on particle production}
Gluons in the wavefunction of the initial onium 
that have a transverse momentum
smaller than the saturation momentum of the nucleus undergo multiple
scatterings with the nucleus and are freed, while the nucleus will essentially
be transparent to gluons with transverse momenta larger than the 
nuclear saturation scale. The {mean} multiplicity of the produced gluons
per unit rapidity measured at central rapidity $y\simeq 0$ is 
related to the distribution $xG(x,Q_s^2(y_0))$ of gluons in the onium 
of lightcone momentum $k_+=x p_+$
integrated up to the transverse momentum scale $Q_s(y_0)$ as follows\footnote{
$x=k_+/p_+=\left[k_+\sqrt{2}/Q_s(y_0)\right]\times\left[Q_s(y_0)/\sqrt{2}p_+\right]
=\sqrt{k_+/k_-}\times Q_s(y_0)/\sqrt{2}p_+$,
where the last equality stems from the mass-shell condition $2k_+k_-=Q_s^2(y_0)$.
The first factor is then the exponential of the rapidity of the measured gluon relative to the rapidity of
the nucleus, namely $e^{y_0}$.

An equivalent formula is $x=e^{-(Y-y_0)}Q_s(y_0)/M$, where $M$ is the mass of the onium.
}
\cite{Kovchegov:1998bi,Mueller:2016xti}:
\be
\left.\frac{dN}{dy}\right|_{y\simeq 0}=xG(x,Q_s^2(y_0))\ ,\ \
\text{where}\ \ x=e^{y_0}\frac{Q_s(y_0)}{\sqrt{2}\,p_+}.
\label{eq:dNdyxG}
\ee\\

Formula~(\ref{eq:dNdyxG}) relates expectation values, namely mean quantities, where
the averages are taken over events.
In a given event,
just before the collision occurs, the incoming onium is found in a particular 
Fock state (essentially made of gluons if its rapidity is large enough).
The number of gluons and their momenta are random variables whose values 
fluctuate from event to event. We are going to assume that an equation similar to
Eq.~(\ref{eq:dNdyxG}) holds as an identity between the {\it random variables}
``number of produced particles per unit rapidity in the particular considered event''
and ``number of gluons in {the corresponding} {realization} of the quantum
evolution of the onium''. Then the {distribution} of the multiplicity of particles
produced is tantamount to the {distribution} of the gluon
number in the particular realization of the partonic content of the
onium at the time of its interaction with the nucleus.

We will choose the frame (namely the rapidity $y_0$ of the nucleus) in such a way
that the onium (which has the rapidity $Y-y_0$) appears as
a dilute object whose state develops through a (linear)
branching process, while the nucleus
is characterized by a large saturation scale $Q_s(y_0)\gg \Lambda_{\text{QCD}}$,
see Fig.~\ref{fig:plotintro}. 
In the next section, we shall review small-$x$ evolution in the linear regime.


\section{\label{sec:background}Background on the small-$x$ evolution in the linear regime}

Throughout, we will use the large number-of-color limit which will
enable us to always represent the partonic content 
of the onium by a set of color dipoles~\cite{Mueller:1993rr}.\\

We consider the set of
dipoles generated by small-$x$ evolution
starting with a dipole of size $x_{01}$, which is our initial condition.
Our goal in this paper amounts to computing
the probability to observe a number $n(r_s;x_{01},y)$ of dipoles of size
larger than $r_s$ at some rapidity $y$ in one event, starting the evolution
from a single dipole.\\

At double logarithmic accuracy\footnote{The most straightforward way to check  Eq.~(\ref{eq:relxGn1}) 
is to compare the explicit expressions its left and right-hand sides.
In the double-log approximation, the gluon density $xG$ reads
$xG(x,Q^2)=\sqrt{\bar\alpha\ln Q^2/y}\,I_1(2\sqrt{\bar\alpha y \ln Q^2})$, while
the formula for $n^{(1)}$ is rederived below, see Eq.~(\ref{eq:n1DL}).
}, there is a simple
relation between the ordinary (namely integrated) gluon density
that appears in Eq.~(\ref{eq:dNdyxG}) and the first moment of $n$:
\be
xG(x,Q_s^2(y_0))=\left.\frac{\partial}{\partial y}n^{(1)}(r_s=1/Q_s(y_0);x_{01},y)\right|_{y=\ln 1/x},
\label{eq:relxGn1}
\ee
where $n^{(1)}$ is the dipole number averaged over the events: 
$n^{(1)}=\langle n\rangle$.
We shall assume that this relation would also hold as 
an identity between the gluon density and the rate of evolution of
the dipole number with the rapidity {\it in each realization} of the quantum evolution.
\\

We start by discussing
the QCD evolution in the dipole model. Then, we establish 
the evolution equations for the dipole number.


\subsection{\label{subsec:dipole}QCD evolution and dipole branching}

The QCD evolution results from the branching of the dipoles when the
longitudinal phase space opens, namely
when the rapidity grows. This branching is due to the emission
of a gluon at position ${\boldsymbol x_2}$ in the transverse plane
from one of the endpoints of the dipoles,
at respective positions ${\boldsymbol x_0}$ and ${\boldsymbol x_1}$. 
The rate of emission, per unit rapidity $dy$ and transverse surface 
$d^2{\boldsymbol x_2}$, is given by the BFKL kernel
\be
K_0({\boldsymbol x_2};{\boldsymbol x_0},{\boldsymbol x_1})
=\frac{\bar\alpha}{2\pi}
\frac{x_{01}^2}{x_{02}^2 x_{12}^2},
\label{eq:K0}
\ee
where $x_{ij}=|{\boldsymbol x_i}-{\boldsymbol x_j}|$ 
is the size of the corresponding dipole.
If one is not interested in the absolute position of the dipoles in the transverse
plane but only in their sizes, then it is useful to express this rate
as a rate of emission per unit size and rapidity:
\be
K(x_{02},x_{12};x_{01})=\bar\alpha
\frac{4x_{01}^2}{x_{02}x_{12}}\frac{1}
{\sqrt{\left[(x_{12}+x_{02})^2-x_{01}^2\right]
\left[x_{01}^2-(x_{12}-x_{02})^2\right]}}.
\label{eq:K}
\ee
This formula holds whenever the distances $x_{01}$, $x_{02}$ and $x_{12}$
may represent the length of the edges of a triangle, and $K=0$ if this
condition is not verified.
$K(x_{02},x_{12};x_{01})dx_{02}dx_{12}dy$ is interpreted as the probability
that a dipole of size $x_{01}$ split to two dipoles of sizes
$x_{02}$ and $x_{12}$ respectively (up to $dx_{02}$ and $dx_{12}$ resp.) when
the rapidity increases by $dy$.\\

Finally, it will prove convenient to introduce logarithmic dipole sizes 
$\rho_{ij}=\ln x_{01}^2/x_{ij}^2$. In this case, the splitting probability reads
$\tilde K(\rho_{02},\rho_{12})d\rho_{02}d\rho_{12}dy$, where
\be
\tilde K(\rho_{02},\rho_{12})
=\frac{\bar\alpha}{\sqrt{
\left[\left(e^{-\rho_{12}/2}+e^{-\rho_{02}/2}\right)^2-1\right]
\left[1-\left(e^{-\rho_{12}/2}-e^{-\rho_{02}/2}\right)^2\right]}}.
\label{eq:Ktilde}
\ee
Note that due to scale invariance, the kernel effectively depends
on two independent real variables only.


\subsection{Dipole number}

Let us introduce the probability $P_n(r_s;x_{01},y)$ of having $n$ dipoles
of size larger than $r_s$ after evolution over the rapidity $y$, starting 
with a single dipole of size $x_{01}$. It is easy to establish an equation
for $P_n$. To this aim, we assume that $P_n(r_s;x_{01},y)$ is known for all $n$ and
all $x_{01}$ and we express $P_n(r_s;x_{01},y+dy)$, merely translating the branching
process described in Sec.~\ref{subsec:dipole} into an equation:
\begin{multline}
P_n(r_s;x_{01},y+dy)=P_n(r_s;x_{01},y)
\left(1-dy\int {dx_{02}dx_{12}} K(x_{02},x_{12};x_{01})\right)\\
+dy \int  {dx_{02}dx_{12}} K(x_{02},x_{12};x_{01}) \sum_{m=1}^{n-1}P_m(r_s;x_{02},y)
P_{n-m}(r_s;x_{12},y).
\end{multline}
Hence
\be
\frac{\partial P_n(r_s;x_{01},y)}{\partial y}=
\int
{dx_{02}dx_{12}} K(x_{02},x_{12};x_{01}) 
\left[
 \sum_{m=1}^{n-1}P_m(r_s;x_{02},y)
P_{n-m}(r_s;x_{12},y)
-P_n(r_s;x_{01},y)
\right].
\ee
This system of equations for the set of $P_n$'s
may be represented by a single equation
for the generating function $Z$ of the factorial moments of the dipole number, which is defined as
\be
Z(r_s;x_{01},y|u)=\sum_{n=1}^\infty u^n P_n(r_s;x_{01},y).
\ee
Indeed, $Z$ is easily seen 
to obey the Balitsky-Kovchegov (BK) equation\footnote{
Strictly speaking, the BK equation was established as an equation for the $S$-matrix element
for the forward elastic scattering of a dipole off a large nucleus.
The evolution of the generating function~$Z$ we discuss here
was first addressed in Ref.~\cite{Mueller:1993rr}. The two evolution equations
can be written in the very same form.}~\cite{Balitsky:1995ub,Kovchegov:1999yj}
\be
\frac{\partial}{\partial y}Z(r_s;x_{01},y|u)=
\int {dx_{02}dx_{12}} K(x_{02},x_{12};x_{01}) 
\left[
Z(r_s;x_{02},y|u)Z(r_s;x_{12},y|u)-Z(r_s;x_{01},y|u)
\right].
\label{eq:BK}
\ee
Since the initial condition is a single dipole of size $x_{01}>r_s$, then
obviously 
\be
Z(r_s;x_{01},y=0|u)=u.
\ee 
We also have 
\be
Z(r_s;x_{01},y|u=1)=1
\ee
from the unitarity relation $\sum_{n=1}^\infty P_n=1$.\\

The factorial moments $n^{(k)}$ of the dipole numbers
are obtained from $Z$ by derivation with respect to the
dummy variable $u$:
\be
n^{(k)}(r_s;x_{01},y)\equiv \langle n(n-1)\cdots (n-k+1)\rangle
=\left.\frac{\partial^k}{\partial u^k}\right|_{u=1}Z(r_s;x_{01},y|u).
\ee
Applying $k$ times the derivation operator to the BK equation~(\ref{eq:BK}),
we see that the set of the factorial moments
$n^{(k)}$ solves a hierarchy of
integro-differential equations:
\begin{multline}
\frac{\partial}{\partial y}n^{(k)}(r_s;x_{01},y)=
\int{dx_{02}dx_{12}} K(x_{02},x_{12};x_{01}) 
\bigg[
n^{(k)}(r_s;x_{02},y)+n^{(k)}(r_s;x_{12},y)-n^{(k)}(r_s;x_{01},y)\\
+\sum_{j=1}^{k-1}\left(\begin{matrix}{k}\\{j}\end{matrix}\right)
n^{(k-j)}(r_s;x_{02},y)n^{(j)}(r_s;x_{12},y)
\bigg].
\label{eq:n(k)}
\end{multline}
The equation for $n^{(1)}$ is the dipole version of the
usual BFKL equation for the mean gluon number. 
Taking into account the initial condition 
$n^{(1)}(r_s;x_{01},y=0)=\Theta(\ln x_{01}^2/r_s^2)$ (single dipole of size $x_{01}$), 
its solution reads\footnote{Note that with our definition of $n^{(1)}(r_s;x_{01},y)$, 
it only depends on the dipole size and not on its orientation nor on 
the position of its center in the transverse plane. The solution to the BFKL equation 
we give here is restricted to zero conformal spin.}
\be
n^{(1)}(r_s;x_{01},y)=\int_{\frac12-i\infty}^{\frac12+i\infty}
\frac{d\gamma}{2i\pi\gamma}
e^{\bar\alpha \chi(\gamma)y} \left(\frac{x_{01}^2}{r_s^2}\right)^{\gamma},
\label{eq:n(1)}
\ee
where $\chi(\gamma)=2\psi(1)-\psi(\gamma)-\psi(1-\gamma)$.\\

The higher moments obey 
an evolution equation which has the same kernel as
the BFKL equation, but with a nontrivial 
source term represented by the inhomogeneous term
in Eq.~(\ref{eq:n(k)}).


\subsection{Integral expression for the higher moments}

It is useful to introduce the number density
$f(x;x_{01},y)$ of dipoles of transverse size $x$ present in
the system after evolution of an initial dipole of size $x_{01}$
over $y$ units of rapidity. We define $f$
in such a way that 
\be
\int_{r_S^2}^{+\infty}\frac{dx^2}{x^2}f(x;x_{01},y)=n(r_s;x_{01},y)
\ee 
is the integrated number of dipoles.
The mean dipole number density $f^{(1)}=\langle f\rangle$ reads
\be
f^{(1)}(x;x_{01},y)=\int_{\frac12-i\infty}^{\frac12+i\infty}
\frac{d\gamma}{2i\pi}
e^{\bar\alpha \chi(\gamma)y} \left(\frac{x_{01}^2}{x^2}\right)^{\gamma}.
\label{eq:f(1)}
\ee

We can readily express $n^{(2)}$ with
the help of $f^{(1)}$ and $n^{(1)}$. 
One easily checks\footnote{Take for instance the derivative
of Eq.~(\ref{eq:f(2)}) with respect to $y$ and
identify its r.h.s. to the r.h.s. of the evolution equation for $n^{(2)}$ 
in Eq.~(\ref{eq:n(k)}) with $k=2$.}
that the following formula holds true:
\begin{multline}
n^{(2)}(r_s;x_{01},y)=2\int \frac{dx_{23}^2}{x_{23}^2}\int_0^y dy_1\,
f^{(1)}(x_{23};x_{01},y_1)\\
\times\int {dx_{24}dx_{34}}K(x_{24},x_{34};x_{23})\,
n^{(1)}(r_s;x_{24},y-y_1)
n^{(1)}(r_s;x_{34},y-y_1).
\label{eq:f(2)}
\end{multline}
As for the higher moments of $n$, similar
equations may be written to express $n^{(k)}$ with the help
of the $n^{(j)}$'s with $j<k$
The existence of such recursion relations is of course
just related to the tree structure of the dipole evolution.


\section{\label{sec:collinear}Collinear limit}

Let us first study the collinear limit
in which $r_s$ is much smaller than the size $x_{01}$
of the initial dipole. 
The dominant contribution to the moments of the dipole number
is given by the configurations in which successive
dipole splittings are strongly ordered in size,
namely $x_{02}\ll x_{01}$ and $x_{12}\simeq x_{01}$ 
or $x_{02}\simeq x_{01}$ and $x_{12}\ll x_{01}$. 
This translates into inequalities for the logarithmic dipole sizes
introduced above in Sec.~\ref{subsec:dipole}, $\rho_{02}>0$ and $\rho_{12}\simeq 0$ 
or $\rho_{02}\simeq 0$ and $\rho_{12}>0$.
Technically, the BFKL kernel boils down to a uniform
distribution in the logarithm of the dipole sizes:
$\tilde K(\rho_{02},\rho_{12})\simeq
\frac{\bar\alpha}{2}\left[\Theta(\rho_{02})\delta(\rho_{12})
+\Theta(\rho_{12})\delta(\rho_{02})\right]$.\\

We write $\rho_s=\ln x_{01}^2/r_s^2$ and use logarithmic variables as the
argument of $Z$ that represents the dipole sizes.
The BK equation~(\ref{eq:BK}) simplifies to
\be
\partial_y Z(\rho_s,y|u)=\bar\alpha Z(\rho_s,y|u)\int_0^{\rho_s}
d\rho^\prime\left[
Z(\rho^\prime,y|u)-1
\right].
\label{eq:BKDL}
\ee
The upper bound on the $\rho^\prime$ integration
implements the fact that a dipole of a given size cannot
split to larger dipoles in the collinear limit.


\subsection{Moments of the dipole number}

In the same manner as in Sec.~\ref{sec:background},
the equations for the factorial moments~$n^{(k)}$ of
the dipole number are easily obtained 
by taking $k$ derivatives of the
BK equation in the collinear limit~(\ref{eq:BKDL})
with respect to $u$, at $u=1$:
\be
\partial_yn^{(k)}(\rho_s,y)-\bar\alpha\int_0^{\rho_s} d\rho^\prime
n^{(k)}(\rho^\prime,y)
=\bar\alpha
\sum_{j=1}^{k-1}
\left(\begin{matrix}{k}\\{j}\end{matrix}\right)
n^{(j)}(\rho_s,y)
\int_0^{\rho_s} d\rho^\prime\,n^{(k-j)}(\rho^\prime,y),
\label{eq:hierarchy}
\ee
namely
\be
\left\lbrace
\begin{split}
\partial_yn^{(1)}(\rho_s,y)-\bar\alpha 
\int_0^{\rho_s} d\rho^\prime n^{(1)}(\rho^\prime,y)&=0,\\
\partial_yn^{(2)}(\rho_s,y)-\bar\alpha 
\int_0^{\rho_s} d\rho^\prime n^{(2)}(\rho^\prime,y)
&=2\bar\alpha n^{(1)}(\rho_s,y)
\int_0^{\rho_s} d\rho^\prime n^{(1)}(\rho^\prime,y)\\
\cdots &
\end{split}
\right.
\ee\\

For the purpose of trying to understand the
properties of the solutions to the hierarchy~(\ref{eq:hierarchy}), 
it is convenient to start over with Eq.~(\ref{eq:BKDL}) and to
rewrite it as a second-order partial differential
equation
\be
\partial_{\rho_s}\partial_y \ln Z(\rho_s,y|u)=\bar\alpha
\left[Z(\rho_s,y|u)-1\right].
\label{eq:gencum}
\ee
The left-hand side is a second 
derivative of the generating function of the factorial
cumulants $n_c^{(k)}$
of the dipole multiplicity, connected to the 
factorial moments $n^{(k)}$ through the relations 
\be
n^{(1)}_c=n^{(1)},\ \ n^{(2)}_c=n^{(2)}-\left[n^{(1)}\right]^2,\ \
n^{(3)}_c=n^{(3)}-3n^{(2)}n^{(1)}+2\left[n^{(1)}\right]^3,\cdots
\label{eq:cumulants}
\ee
Introducing further the variables $z=2\sqrt{\bar\alpha y\rho_s}$
and $\tilde z=\frac12\sqrt{\rho_s/\bar\alpha y}$,
the differential operator becomes
$\partial_{\rho_s}\partial_y
=\bar\alpha\left(\partial_z^2+\frac{1}{z}\partial_z
-\frac{\tilde z^2}{z^2}\partial_{\tilde z}^2-\frac{\tilde z}{z^2}
\partial_{\tilde z}\right)$.
We then see that it is consistent to look for solutions 
which are independent of the variable $\tilde z$:
We will focus on such solutions in what follows.
To obtain the differential equations
in terms of the $z$ variable,
it is convenient to start from Eq.~(\ref{eq:gencum})
and take again $k$ derivatives with respect to $u$
\be
\left\lbrace
\begin{split}
\left(z^2\partial_z^2+z\partial_z -z^2\right)
n^{(1)}_c(z)
&=0,\\
\left(z^2\partial_z^2+z\partial_z -z^2\right)
n^{(2)}_c(z)
&=z^2 \left[n^{(1)}(z)\right]^2,\\
\left(z^2\partial_z^2+z\partial_z -z^2\right)
n^{(3)}_c(z)
&=z^2\left\{3n^{(2)}(z)n^{(1)}(z)-2\left[n^{(1)}(z)\right]^2\right\},\\
\cdots &
\end{split}
\right.
\ee
We note that the operator appearing in the homogeneous part of these equations is 
the kernel of a Bessel equation.
The solution to the hierarchy cannot be expressed fully analytically,
however, the large-$z$ asymptotics are simple and 
partially known.
Indeed, the same kind of equations appear in the context 
of jet physics~\cite{Dokshitzer:1982ia,Dokshitzer:1991wu}.
Let us nevertheless discuss these asymptotics in some detail.\\

The solution of the equation for the first cumulant (or moment) 
$n^{(1)}_c=n^{(1)}$ with the initial
condition $n^{(1)}(\rho_s,y=0)=1$ is a modified Bessel function of the first kind:
\be
n^{(1)}(z)=I_0(z).
\label{eq:n1DL}
\ee
The first moment $n^{(1)}$ is the mean dipole number. Another
notation for $n^{(1)}$ that we shall use in what follows
is $\bar n$.\\

The next equations in the hierarchy are seen to 
exhibit the same kernel as the equation for~$n^{(1)}$: 
Only the inhomogeneous term differs.
As for $n_c^{(2)}$, we find an exact expression which, once
re-expressed in terms of moments through Eq.~(\ref{eq:cumulants}), reads
\be
n^{(2)}(z)=I_0^2(z)+I_0(z)\int_0^z dz^\prime z^\prime
K_0(z^\prime)I_0^2(z^\prime)-K_0(z)\int_0^z dz^\prime z^\prime
I_0^3(z^\prime).
\ee
The above integrals do not have a simpler expression, however,
we may obtain the large-$z$ expansion of $n^{(2)}$ from
the expansion of the Bessel functions:
\be
I_0(z)\underset{z\rightarrow\infty}{=}
\frac{e^z}{\sqrt{2\pi z}}
\left(1+\frac{1}{8z}+\cdots\right),\qquad
K_0(z)\underset{z\rightarrow\infty}{=}
\sqrt{\frac{\pi}{2z}}
{e^{-z}}
\left(1-\frac{1}{8z}+\cdots\right)
\ee
To first order in $1/z$ and switching back to the $(\rho_s,y)$ variables, 
we get
\be
\frac{n^{(2)}(\rho_s,y)}{\left[n^{(1)}(\rho_s,y)\right]^2}=\frac43
\left(1+\frac{1}{12\sqrt{\bar\alpha y\rho_s}}+O(1/\bar\alpha y\rho_s)\right).
\ee
We may repeat this procedure for the higher moments.
For example we find for $n^{(3)}$:
\be
\frac{n^{(3)}(\rho_s,y)}{\left[n^{(1)}(\rho_s,y)\right]^3}=\frac94
\left(1+\frac{5}{12\sqrt{\bar\alpha y\rho_s}}
+O(1/\bar\alpha y\rho_s)\right)
\ee
The generic structure is
\be
\frac{n^{(k)}(\rho_s,y)}{\left[n^{(1)}(\rho_s,y)\right]^k}
=C_k+O(1/\sqrt{\bar\alpha y\rho_s}),
\ee
where the coefficients $C_k$ are constants.\\

The coefficients $C_k$ may be computed
by inserting the {\it Ansatz} $n^{(k)}(z)=C_k\left[\bar n(z)\right]^k$
into the hierarchy~(\ref{eq:hierarchy}), and by recalling that 
$n^{(1)}(\rho_s,y)\simeq e^{2\sqrt{\bar\alpha\rho_s y}}$.
The $C_k$ are then seen to obey the following recursion~\cite{Dokshitzer:1982ia}:
\be
C_{k\geq 2}=
\frac{k}{k^2-1}\sum_{j=1}^{k-1}\left(\begin{matrix}{k}\\{j}\end{matrix}\right)
\frac{C_j C_{k-j}}{k-j}\ ,\ \
C_1=1.
\label{eq:recursionCk}
\ee
For large $k$, $C_k$ converges\footnote{
It is straightforward to check this asymptotic form 
for the solution by inserting it into Eq.~(\ref{eq:recursionCk}).
The value of the constant $c_\text{DL}$ is obtained numerically.} fastly to 
$C_k\simeq 2k\, k!\, (c_{\text{DL}})^k$, where $c_{\text{DL}}=0.391\cdots$.


\subsection{Multiplicity distribution}

From the knowledge of the large-$k$ and large-$z$
asymptotics of the factorial moments (and hence of the ordinary moments,
due to the exponential increase of $n^{(k)}(z)$ at large $z$), one
can infer the large-$n$ behavior of the distribution of the dipole
multiplicity~$n$.
Indeed,
\be
n^{(k)}(z)=\sum_{n=1}^\infty n^k P_n(z)
\simeq \int_0^{+\infty} dn\,n^k P_n(z)
\label{eq:reln(k)Pn}
\ee
where the second approximate equality holds for large $k$, since the sum
over $n$ is dominated by large values of $n$ in that limit.\\

We continue analytically the moment index $k$ to complex values,
and invert this equation as
\be
P_n(z)=\int\frac{dk}{2i\pi}n^{-k-1}n^{(k)}(z)
\label{eq:relPnn(k)}
\ee
which when we specialize to the collinear limit reads
\be
P_n^\text{DL}(z)=\frac{2}{c_\text{DL}\bar n_\text{DL}(z)}
\int\frac{dk}{2i\pi}
k\,\Gamma(k+1)
\left(\frac{c_\text{DL}\bar n_\text{DL}(z)}{n}\right)^{k+1}.
\ee
A straightforward calculation leads to the final result
\be
P^\text{DL}_n\propto\frac{2}{c_{\text{DL}}\bar n_\text{DL}}
\left(\frac{1}{c_{\text{DL}}}\frac{n}{\bar n_\text{DL}}-1\right)
\exp\left(-\frac{1}{c_{\text{DL}}}\frac{n}{\bar n_\text{DL}}\right).
\label{eq:PnDL}
\ee
Note that the probability distribution
$P_n^\text{DL}$ exhibits Koba-Nielsen-Olesen (KNO) 
scaling~\cite{Koba:1972ng}, namely 
$\bar n_\text{DL}P_n^\text{DL}$ is a function of $n/\bar n_\text{DL}$
only.\\

One may think that the qualitative properties of this distribution
would be kept when one gives up the strong ordering
of the transverse momenta.
This is actually not at all the case, as we will demonstrate
in the next section.


\section{\label{sec:BFKL}Full BFKL evolution}

When the ordering condition between~$r_s$ and $x_{01}$ is released,
then the full BK equation has to be solved.
The difference between the full BK equation~(\ref{eq:BK})
and its DL approximation~(\ref{eq:BKDL}) is that the former
allows splittings to {larger} dipoles, while the latter describes 
only splittings to smaller dipoles.\\

We shall now review the asymptotics of the solution, which were
first derived in Ref.~\cite{Salam:1995zd}, and interpret them.
The most convenient is to analyze the equations for the moments.
Let us start by discussing the second-order factorial moment $n^{(2)}$.


\subsection{Second-order moment}

The second-order factorial moment is obtained by inserting the expressions
of the first moments~(\ref{eq:n(1)}),(\ref{eq:f(1)}) into Eq.~(\ref{eq:f(2)}).
Using the logarithmic variables 
$\rho_s=\ln x_{01}^2/r_s^2$ and $\rho_1=\ln x_{01}^2/x_{23}^2$,
$n^{(2)}$ can be cast as
\be
n^{(2)}(\rho_s,y)=2\int_{-\infty}^{+\infty}d\rho_1\int_0^y dy_1\int 
\frac{d\gamma}{2i\pi}
\frac{d\gamma_1}{2i\pi\gamma_1}
\frac{d\gamma_2}{2i\pi\gamma_2}
P_3(\gamma_1,\gamma_2)
e^{
{\cal E}_2(\gamma,\gamma_1,\gamma_2,y_1,\rho_1;\rho_s,y)
},
\label{eq:n(2)}
\ee
where
\be
{\cal E}_2=
\bar\alpha y_1\chi(\gamma)+\gamma\rho_1
+\bar\alpha(y-y_1)[\chi(\gamma_1)+\chi(\gamma_2)]
+(\gamma_1+\gamma_2)(\rho_s-\rho_1)
\label{eq:E2}
\ee
and
\be
P_3(\gamma_1,\gamma_2)=\int d\delta_1d\delta_2\tilde K(\delta_1,\delta_2)
e^{-\gamma_1\delta_1-\gamma_2\delta_2}
\ee
($\delta_1=\ln x_{23}^2/x_{24}^2$ and $\delta_2=\ln x_{23}^2/x_{34}^2$).
We want to evaluate $n^{(2)}$ in the limit of large
$\rho_s$ and large~$y$.\\

We start by looking for a saddle point in the $\gamma$, $\gamma_1$ 
and $\gamma_2$ variables independently, $\rho_1$ and $y_1$
being fixed for the time being.
We require that the partial derivative
of ${\cal E}_2$ with respect to these variables vanish, which leads to
the unique solution
\be
\chi^\prime(\gc[]{2})=-\frac{\rho_1}{\bar\alpha y_1}\ ,\ \
\chi^\prime(\gc[1]{2})=-\frac{\rho_s-\rho_1}{\bar\alpha(y-y_1)}\ \ \text{and}
\ \ \gc[2]{2}=\gc[1]{2}.
\label{eq:sp1}
\ee
Performing this integral by the steepest-descent method, the pair multiplicity reads
\be
n^{(2)}(\rho_s,y)\simeq \frac{2}{(2\pi)^{3/2}}
\int_{-\infty}^{+\infty}d\rho_1\int_0^y dy_1
\frac{P_3(\gc[1]{2},\gc[2]{2})}{\gc[1]{2}\gc[2]{2}}
\times
{\frac{1}{\sqrt{|\det H|}}}
e^{
{\cal E}_2(\gc[]{2},\gc[1]{2},\gc[2]{2},y_1,\rho_1;\rho_s,y)
},
\ee
where $H$ is the matrix of the second derivatives of ${\cal E}_2$ 
evaluated at the saddle point.\\

The saddle point $(\gc[]{2},\gc[1]{2}=\gc[2]{2})$
depends on $y_1$ and on $\rho_1$, over which we still need to integrate.
We are going to search for a stationary point of 
${\cal E}_2$ in the two remaining integration variables $y_1$ and $\rho_1$.
The partial derivatives of ${\cal E}_2$ with respect to $\rho_1$ and $y_1$ 
taken at the saddle point read
\begin{subequations}
\begin{align}
\label{eq:dEdr1}
\frac{\partial{\cal E}_2}{\partial\rho_1}&=\gc[]{2}-2\gc[1]{2}\\
\frac{\partial{\cal E}_2}{\partial y_1}&=
\bar\alpha\left[\chi(\gc[]{2})-2\chi(\gc[1]{2})\right].
\label{eq:dEdy1}
\end{align}
\label{eq:dE}
\end{subequations}

\subsubsection{Global saddle point}

\begin{figure}
\begin{center}
\includegraphics[width=0.49\textwidth]{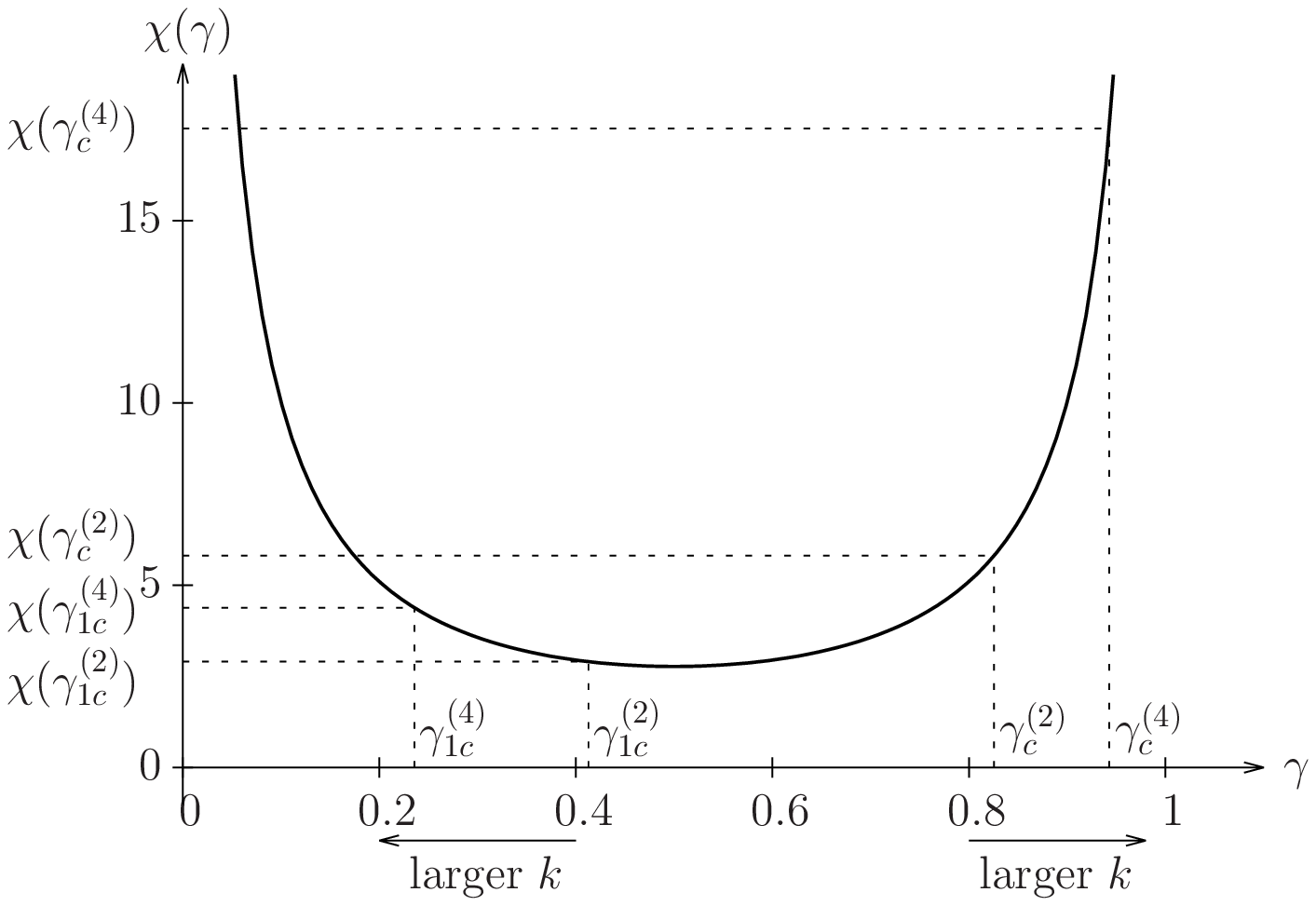}
\includegraphics[width=0.49\textwidth]{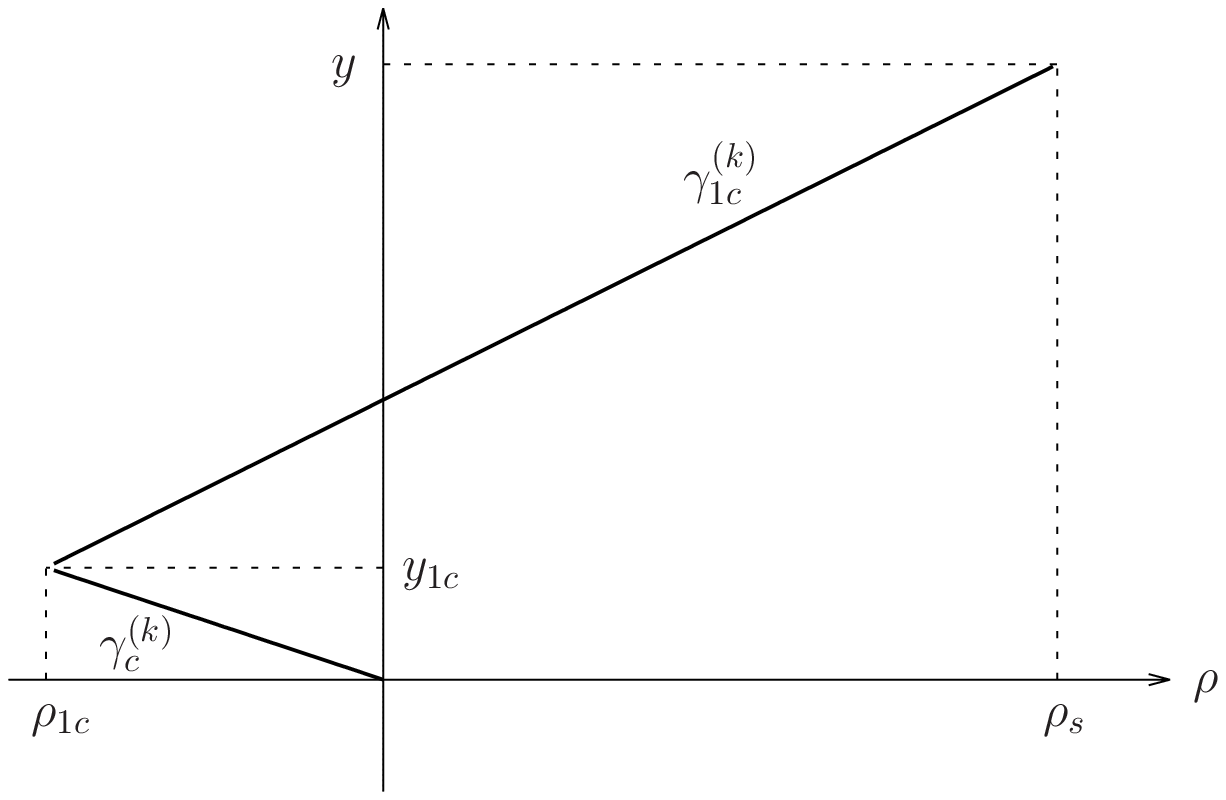}
\end{center}
\caption{\label{fig:plotchi2}
\small
{\it Left}: Characteristic function
of the BFKL kernel and the global saddle point solutions for
the second and the fourth moments. One sees that the higher the moment index,
the more the anomalous dimensions corresponding to the
first and the second steps of the evolution get attracted by the collinear
and anticollinear singularities at $\gamma=1$ and $\gamma=0$ respectively.
{\it Right}: Schematic representation of the evolution
path selected by the moments $n^{(k)}$ in the $(\rho,y)$ plane.
}
\end{figure}

For a global saddle point of the multiple integral
to exist, the derivatives 
in Eq.~(\ref{eq:dE})
must vanish simultaneously, which requires
\be
\gc[1]{2}=\gc[2]{2}=\frac{\gc[]{2}}{2}\ ,\ \
\chi(\gc[]{2})=2\chi(\gc[1]{2}).
\label{eq:sp2}
\ee
Eq.~(\ref{eq:sp2}) can be solved numerically: $\gc[1]{2}=0.412796\cdots$.
The solution is represented in Fig.~\ref{fig:plotchi2}.
We note that $\chi^\prime(\gc[1]{2})<0$ and $\chi^\prime(\gc[]{2})>0$.
Once the value of $\gc[1]{2}$ is fixed, Eq.~(\ref{eq:sp1}) implies
\be
\bar\alpha y_{1c}=
\frac{\chi^\prime(\gc[1]{2})\bar\alpha y+{\rho_s}}
{\chi^\prime(\gc[1]{2})-{\chi^\prime(\gc[]{2})}}\ ,\ \
\rho_{1c}=-\chi^\prime(\gc[]{2})\bar\alpha y_{1c}.
\label{eq:y1crho1c}
\ee
We note that this solution requires $\rho_{1c}$ to be negative.
Conversely, if $\rho_{1c}<0$, then this saddle point solution exists
provided $0<y_{1c}<y$, a condition that is satisfied whenever
the external parameters obey the ordering relation
\be
\rho_s<-\chi^\prime(\gc[1]{2})\bar\alpha y,
\label{eq:spcondition}
\ee
which follows from the first equation in~(\ref{eq:y1crho1c}), 
conveniently rewritten as
\be
\rho_s=-\chi^\prime(\gc[1]{2})\bar\alpha y
+\bar\alpha y_{1c}\left[\chi^\prime(\gc[1]{2})-\chi^\prime(\gc[]{2})\right]
\ee
and from the negativity of the second term in the r.h.s.\\

The physical picture of this solution is the following:
Starting from the dipole of size $x_{01}$, the first part of the evolution,
which typically takes place over the first 
$\bar\alpha y_{1c}$ units of rapidity, produces a larger dipole
(since $\rho_{1c}$ is negative).
The latter decays to two dipoles, which subsequently evolve independently
over the rapidity range $y-y_{1c}$.
The saddle-point solution is represented on the graph of the $\chi$-function
in Fig.~\ref{fig:plotchi2}, together with the evolution path
in the $(\rho,y)$ plane.

\subsubsection{Connection with the DL limit}

If the global saddle-point solution does not exist, then
the values of $\rho_{1}$ 
that contribute to the integration in Eq.~(\ref{eq:n(2)})
are essentially positive.
In this case, the saddle point equations~(\ref{eq:sp1}) 
impose that both $\gc[1]{2}$ and $\gc[]{2}$ be less than $\frac12$.
We can require again the stationarity of
${\cal E}_2$ with respect to the variations of $\rho$,
which, using (\ref{eq:dEdr1}), leads to
the condition $\gc[]{2}=2\gc[1]{2}$. 
Hence $2\gc[1]{2}=\gc[]{2}<\frac12$, which
trivially implies $\chi(\gc[]{2})<2\chi(\gc[1]{2})$.
Then, according to Eq.~(\ref{eq:dEdy1})
the integral is strongly dominated by the region $y_1\ll y$. 
Indeed, an upper bound for the values of
$\bar\alpha y_1$ which contribute significantly
to the integral is\footnote{
This bound may be derived 
from the form of the $y$-dependence of ${\cal E}_2$
(see Eq.~(\ref{eq:dEdy1})) and
from the convexity properties
of the function 
$2\chi(\gamma)-\chi(2\gamma)$.
}
$1/[2\chi(\frac14)-\chi(\frac12)]=0.18\cdots$,
a number which is small compared to $\bar\alpha y$.\\

\begin{figure}
\begin{center}
\includegraphics[width=0.49\textwidth]{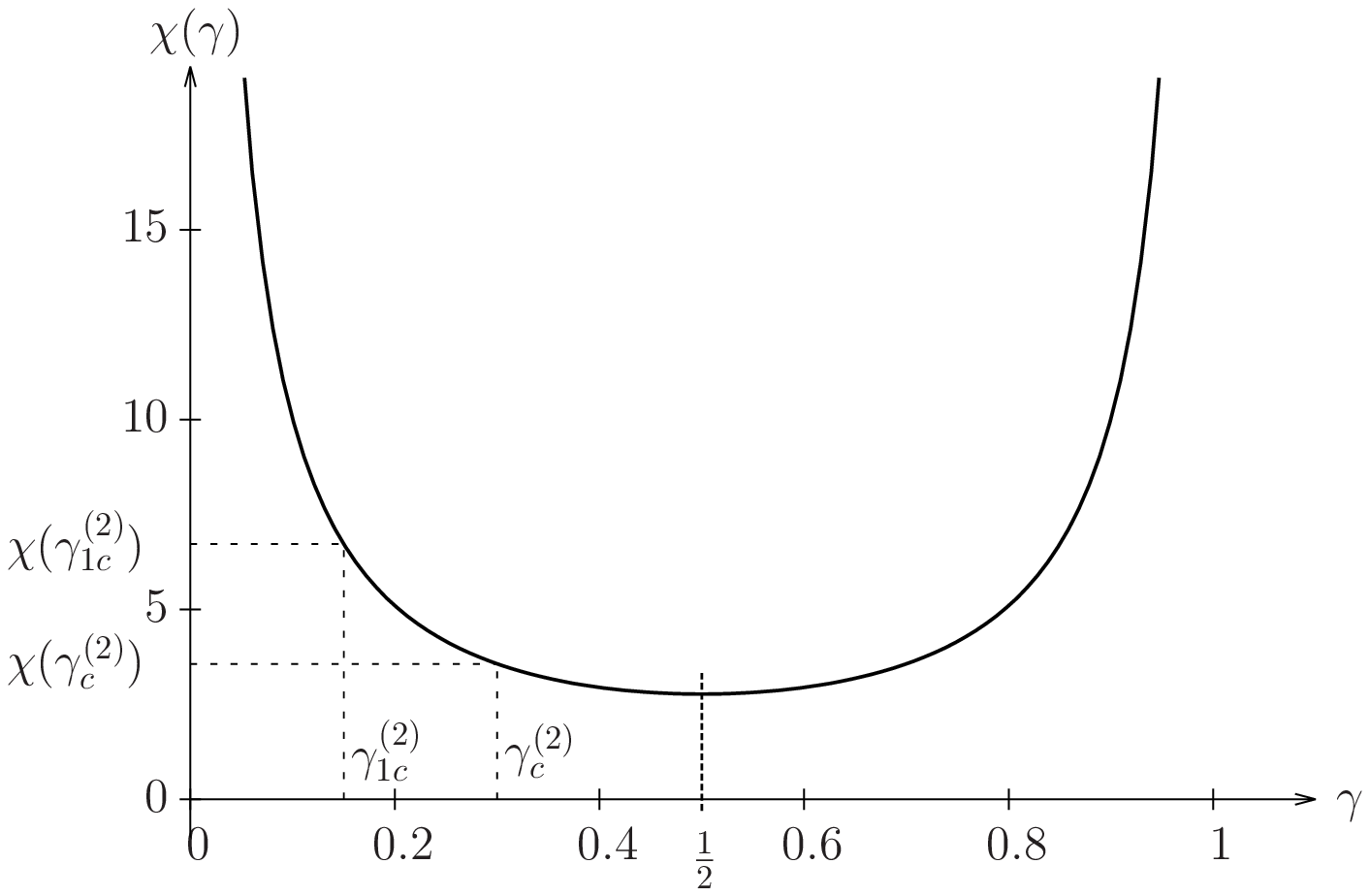}
\includegraphics[width=0.49\textwidth]{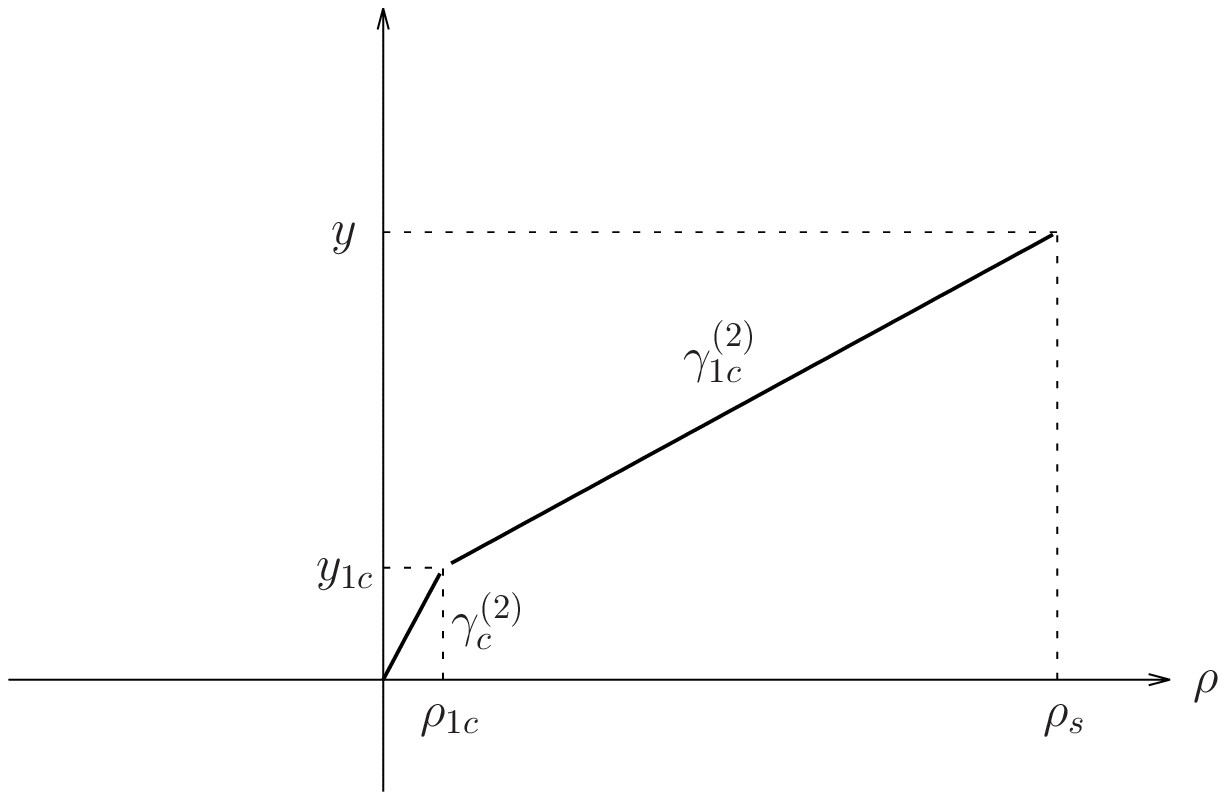}
\end{center}
\caption{\label{fig:plotchi2no}
\small {\it Left}: Solution for
the second moment in the case in which
there is no global saddle point.
$\gc[]{2}$ and $\gc[1]{2}$ are not completely fixed in
this case: They however satisfy the relation
$2\gc[1]{2}=\gc[]{2}<\frac12$.
{\it Right}: Schematic representation of the evolution
path selected by the moment $n^{(2)}$ in the $(\rho,y)$ plane.
There is no step backward. This kind of path is favored
when $\bar\alpha y$ is not large compared to $\rho_s$.
}
\end{figure}

Note that $\chi^\prime(\gc[]{2})>\chi^\prime(\gc[1]{2})$: 
From the saddle-point equation~(\ref{eq:sp1}),
we see that the relevant integration region for $\rho_1$
is $0<\rho_1<\rho_s y_1/y\ll \rho_s$.
Hence the evolution that leads to a pair of dipoles
just consists in two independent evolutions of single dipoles starting
almost right from the beginning of the branching process.
In this regime, the collinear limit studied in Sec.~\ref{sec:collinear}
is relevant throughout the whole evolution.
The solution in this case is represented in Fig.~\ref{fig:plotchi2no}.
\\

The solutions we have just found are qualitatively different
from what one would find in the case of a branching-diffusion process.
This question is addressed in some detail 
in Appendix~\ref{app:branchingdiffusion}.
The dipole model can be identified with 
such a process when one observes the dipole density
at a fixed impact parameter, but when one integrates over the impact parameter
as we do here, then the hotspots generated by the
collinear singularities dominate, and the dipole
branching is no longer a diffusion in the logarithmic dipole sizes.\\

We are now going to investigate in the same spirit 
the higher moments of the dipole multiplicity.


\subsection{Higher moments}

The integral expression of the moment of order $k$ is a complicated
multiple integral of products of mean dipole numbers and densities.
For its complexity, we will not be able to study in detail
its limits in the same way as for $n^{(2)}$.
Therefore, we will assume that the global saddle-point solution
essentially consists in two steps. We will see that
for large enough values of $k$, 
the global saddle point always exists.
The first step is then the production of a larger dipole
and the second its decay into much smaller dipoles.\\

Let us write directly the saddle-point solution:
\be
n^{(k)}(\rho_s,y)\propto e^{{\cal E}_k(\gc[]{k},\gc[1]{k},\cdots,\gc[k]{k};
y_{1c},\rho_{1c};\rho_s,y)},
\ee
where
\be
{\cal E}_{k}
=\bar\alpha y_{1c}\chi(\gc[]{k})
+\gc[]{k}\rho_{1c}
+{\bar{\alpha}}(y-y_{1c}) \left[\chi(\gc[1]{k})+\cdots+\chi(\gc[k]{k})\right]
+(\gc[1]{k}+\cdots+\gc[k]{k})(\rho_s-\rho_{1c})
\ee
generalizes Eq.~(\ref{eq:E2}) taken at the saddle point,
and the following relations must hold true:
\be
\gc[1]{k}=\cdots=\gc[k]{k}=\frac{\gc[]{k}}{k}\ ,\ \
\chi(\gc[]{k})=k\chi(\gc[1]{k})\ ,\ \
\chi^\prime(\gc[]{k})=-\frac{\rho_1}{\bar\alpha y_1}\ ,\ \
\chi^\prime(\gc[1]{k})=-\frac{\rho_s-\rho_1}{\bar\alpha(y-y_1)}.
\ee
For large $k$, according to the first relation,
necessarily $\gc[1]{k}\rightarrow 0$. The second relation imposes
$\gc[]{k}\rightarrow 1$. Hence one may replace the complete
expression of $\chi(\gamma)$ by its collinear $\chi(\gamma)\simeq 1/\gamma$
(resp. anticollinear $\chi(\gamma)\simeq 1/(1-\gamma)$) limit whenever this
function or its derivative are evaluated at $\gamma=\gc[1]{k}$
(resp. $\gamma=\gc[]{k}$).
The saddle-point equations lead to $\gc[]{k}\simeq 1-1/k^2$,
$\gc[1]{k}\simeq 1/k$ at large $k$.
Then 
\be
\bar\alpha y_{1c}\simeq \bar\alpha y/k^2\ \ \text{and}\ \ 
\rho_{1c}\simeq -k^2 \bar\alpha y. 
\ee
The saddle-point solution should exist as soon as 
$k>\sqrt{\rho_s/\bar\alpha y}$: Hence for large enough~$k$,
it will always be the only relevant solution.\\

Using this determination of the saddle-point parameters, we find
${\cal E}_k\simeq \rho_s+k^2\bar\alpha y$. Hence
the large-$k$ expression for the $k$-th moment of the dipole number
is
\be
n^{(k)}(\rho_s,y)\propto e^{\rho_s+k^2\bar\alpha y}.
\label{eq:n(k)approx}
\ee\\

The interpretation of this solution is straightforward.
The $k$ dipoles that are measured at rapidity $y$
stem from the evolution of a ``common
ancestor'', which was produced at rapidity 
$y_{1c}\simeq y/k^2\rightarrow 0$ (at large
$k$). This ancestor dipole has a size which is of the order
of $e^{k^2\bar\alpha y}$ times bigger than the initial dipole.\\

We have actually just recovered the picture that was argued for in 
Ref.~\cite{Salam:1995zd}, where the backward step of the evolution
was assumed to coincide with the very first splitting, 
and the further evolution
was replaced by its collinear limit.


\subsection{Multiplicity distribution and physical picture}

From the expression of the moments of the dipole multiplicity
of order $k$ in the large-$k$ limit, we can go back to the distribution
using the relation~(\ref{eq:relPnn(k)})
and replacing therein $n^{(k)}$ by the expression in
Eq.~(\ref{eq:n(k)approx}).
The integral to perform is Gaussian. The result reads \cite{Salam:1995zd}
\be
P_n(r_s;x_{01},y)\propto \frac{x_{01}^2}{r_s^2}
\frac{1}{n}\exp
\left(
-\frac{\ln^2 n}{4\bar\alpha y}
\right).
\label{eq:distribBFKL}
\ee
The large-$n$ tail of this distribution is much fatter than
the one of the DL limit~(\ref{eq:PnDL}).
Moreover, this probability distribution does not obey KNO scaling.\\

If one focusses on larger multiplicities, then 
the evolution goes through large-size dipoles, much larger than
the initial dipole. If the latter models a hadron of typical size 
$1/\Lambda_\text{QCD}$,
the production of much larger dipoles should be
cut off by confinement. 
We shall now introduce a
qualitative model for these effects, and propose a solution.


\section{\label{sec:BFKLconf}Evolution in the presence of confinement}

Confinement is expected to act as a cutoff that prevents
dipoles of size larger
than typically $R\sim 1/\Lambda_{\text{QCD}}$ to be created.
Here we are going to pick a simple model that enable us
to arrive at analytical results.\\

To this aim, we go back to the equation~(\ref{eq:n(k)}) for the moments
of the dipole number
and implement a Gaussian cutoff on the size of the produced
dipoles.
For technical reasons, it is convenient to use the form of the
kernel in Eq.~(\ref{eq:K0}) and to enforce the cutoff
through the substitution
\be
K_0({\boldsymbol x_2};{\boldsymbol x_0},
{\boldsymbol x_1})\rightarrow
\frac{\bar\alpha}{2\pi}
\frac{x_{01}^2}{x_{02}^2 x_{12}^2}e^{-(x_{02}^2+x_{12}^2)/(2R^2)}
\ee
in such a way that Eq.~(\ref{eq:n(k)}) be replaced by
\begin{multline}
\frac{\partial}{\partial y}n^{(k)}(r_s;x_{01},y)=
\bar\alpha\int\frac{d^2 {\boldsymbol x_2}}{2\pi}
\frac{x_{01}^2}{x_{02}^2 x_{12}^2}e^{-(x_{02}^2+x_{12}^2)/(2R^2)}
\bigg[
n^{(k)}(r_s;x_{02},y)+n^{(k)}(r_s;x_{12},y)\\
-n^{(k)}(r_s;x_{01},y)
+\sum_{j=1}^{k-1}\left(\begin{matrix}{k}\\{j}\end{matrix}\right)
n^{(k-j)}(r_s;x_{02},y)n^{(j)}(r_s;x_{12},y)
\bigg].
\label{eq:n(k)cutoff}
\end{multline}
With respect to the case without cutoff studied in the
previous section, only dipoles of size less than $R$
may be created with high probability. 
Hence the larger span for the rapidity evolution
is between the dipole sizes $R$ and $r_s$.
It is then natural to try an {\it Ansatz} of the form
\be
n^{(k)}(r_s;x_{01},y)=\frac{x_{01}^2}{R^2}e^{-x_{01}^2/(2R^2)}C_k
\left[n^{(1)}(r_s;R,y)\right]^k
\ee
for large $k$.
We are going to check that this form is indeed an asymptotic
solution.\\

As in the case of $n^{(1)}$, we may assume that the integral in Eq.~(\ref{eq:n(1)})
is dominated by the value of the integrand at the saddle point $\gamma_s$.
Discarding the prefactors, we write
\be
n^{(1)}(r_s;R,y)\sim \left(\frac{R^2}{r_s^2}\right)^{\gamma_s}
e^{\bar\alpha y \chi(\gamma_s)}
\ee
and $\gamma_s$ solves $\bar\alpha y\chi^\prime(\gamma_s)+\ln R^2/r_s^2=0$.
Looking at Eq.~(\ref{eq:n(k)cutoff}), we see that the nonlinear
terms dominate the large-$y$ solution because of the
$y$-dependence of our {\it Ansatz} for $n^{(k)}$.
Inserting the latter into Eq.~(\ref{eq:n(k)cutoff}) deprived of the
linear terms in the right-hand side, the integro-differential
equation for $n^{(k)}$ becomes a recursion for the constants $C_k$
\be
e^{-x_{01}^2/(2R^2)}k C_k=\frac{1}{\chi(\gamma_s)}
\sum_{j=1}^{k-1}\left(\begin{matrix}{k}\\{j}\end{matrix}\right)
\int \frac{d^2{\boldsymbol x_2}}{2\pi R^2}e^{-(x_{02}^2+x_{12}^2)/R^2}C_j C_{k-j}
\ee
which is easily simplified to
\be
C_k=\frac{1}{4\chi(\gamma_s)}\frac{1}{k}
\sum_{j=1}^{k-1}\left(\begin{matrix}{k}\\{j}\end{matrix}\right)
C_j C_{k-j}.
\label{eq:recursionBFKL}
\ee
For large $k$, this recursion is solved by
$C_k\simeq 4\chi(\gamma_s)k!c^k$, where the constant $c$
depends on $C_0$ on which we have no control since the linear term
we dropped in the equation~(\ref{eq:n(k)cutoff}) 
for the moments plays a central role
for the moments of low~$k$.\\

From the large-$k$ moments, one can infer the large-$n$ 
behavior of the probability distribution. We find
\be
P_n(r_s;x_{01},y)\propto 4\chi(\gamma_s)\frac{x_{01}^2}{R^2}e^{-x_{01}^2/(2R^2)}
\frac{1}{c\bar n}e^{-n/(c\bar n)}.
\label{eq:multi}
\ee
Note that the value of the constant $c$ depends on the form of the 
recursion~(\ref{eq:recursionBFKL}), which in turn may depend 
on the details of the infrared cutoff.
Therefore we do not expect this constant to be universal.
However, the exponential form\footnote{A decreasing exponential is also the dominant behavior of the tail of
the negative binomial distribution found in the glasma model, see Refs.~\cite{Lappi:2006fp,Gelis:2009wh}.} 
is very likely to be universal, because it
requires no more than the existence of an infrared absorptive
boundary in the evolution.
Numerical checks would be useful in order to confirm the solution
in Eq.~(\ref{eq:multi}) and to probe its universality.


\section{Summary and outlook}

In this paper, we studied the distribution of
the multiplicity of the particles produced in 
high-energy onium-nucleus collisions in the forward rapidity
region of the onium,
as the simplest possible model for proton-nucleus collisions.\\

We investigated different kinematical regimes. First, we observed
that in the large-rapidity and small-size regime
in which the double-logarithmic (DL) approximation makes sense,
the problem formally maps to jet evolution.
We then showed that when the DL approximation is released,
the large-multiplicity distribution becomes much fatter, due
to the production of very large dipoles in the course of the
evolution. However, confinement does not allow for such
evolution paths: Once the latter are cut off, the tail of
the multiplicity distribution becomes a decreasing exponential.
The analytical results we obtained in the different regimes of
the dipole evolution are shown in Fig.~\ref{fig:plotmc}
and compared to Monte Carlo simulations of a simple one-dimensional toy model
described in Appendix~\ref{app:num}.\\

\begin{figure}[h]
\begin{center}
\includegraphics[width=0.9\textwidth]{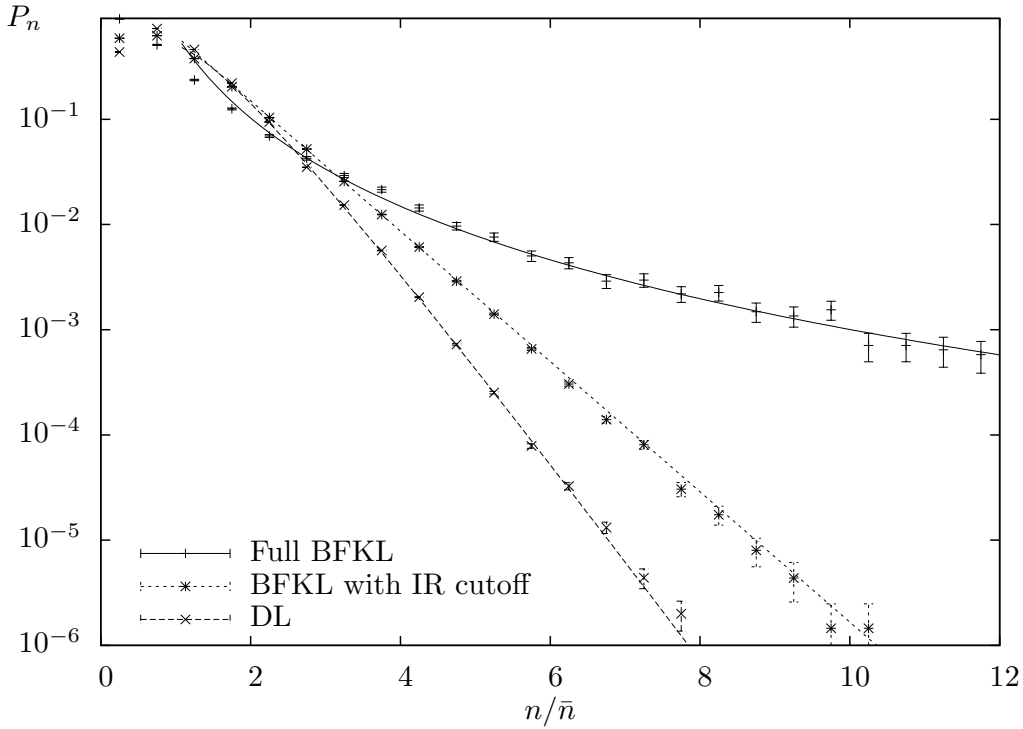}
\end{center}
\caption{\label{fig:plotmc}
\small Distribution of the fluctuations of the multiplicity around
its expected value $\bar n$ from a Monte Carlo simulation
of a simplified model (see Appendix~\ref{app:num}).
The lines represent fits of the analytical 
formulas~(\ref{eq:distribBFKL}),~(\ref{eq:multi}) 
and~(\ref{eq:PnDL})
(from top to bottom).
}
\end{figure}

A natural extension of this work would be to try and build a
more realistic model for proton-nucleus collisions beyond the dipole
model, in order to allow for sensible
comparisons with LHC experimental data. 
From a more theoretical perspective, 
if our assumption on the relation
between event-by-event partonic fluctuations in the
initial state and final-state multiplicity fluctuations
is correct, then the tails of the multiplicity distribution
stem from dense states of gluons present in the wavefunction
of the initially dilute projectile (the onium, or the proton).
Understanding these rare events may open a new window
on very high-density quantum states.

\subsection*{Acknowledgments}

We acknowledge support from the France-US binational exchange program
PICS, project title:
``Fundamental properties of quantum chromodynamics at high energy and high density 
and applications at the LHC''.
T.L. benefited from a Doctoral Mobility Grant 
awarded by the Direction des Relations Internationales
of \'Ecole Polytechnique at the time when this project was initiated.
A.H.M. acknowledges support from the Office of Science of the
U.S. Department of Energy, Grant No.~DE-FG02-92ER40699.
We thank Dr. Cyrille Marquet for suggesting references.


\appendix

\section{\label{app:branchingdiffusion}
Dipole pair multiplicity in a branching-diffusion model}

In this appendix, we shall assume that $\chi(\gamma)$ is a second-order 
polynomial:
\be
\chi(\gamma)=\chi_0+(\gamma-\gamma_0)\chi^\prime_0
+\frac12(\gamma-\gamma_0)^2\chi^{\prime\prime}_0,
\label{eq:diffkernel}
\ee
where $0<\gamma_0<1$ and $\chi_0$, $\chi^\prime_0$, and
$\chi^{\prime\prime}_0>0$
are constants. We assume that $\chi(\gamma)>0$ for all $\gamma$.
Such a second-order polynomial in $\gamma$ represents
the eigenvalues of a branching-diffusion kernel
with a drift.
The technical advantage of such a diffusive model is that the 
integrations over the sizes being Gaussian, they can be performed exactly.\\

 If $\gamma_0=0$, $\chi_0=1$, $\chi^\prime_0=0$ and $\chi^{\prime\prime}_0=2$, 
this is plain branching Brownian motion.
(For a detailed numerical and analytical study of the pair multiplicity in
the branching Brownian motion, see Ref.~\cite{Doche}).
If instead the constants $\chi_0$, $\chi_0^\prime$ and  $\chi_0^{\prime\prime}$
coincide with the values of the functions
$\chi$,
 $\chi^\prime$ and  $\chi^{\prime\prime}$ respectively
evaluated at $\gamma_0$,
with $0<\gamma_0<1$,
then Eq.~(\ref{eq:diffkernel}) defines the so-called diffusive approximation
to the BFKL kernel, which has been studied in QCD. 
The associated stochastic process is a branching random walk
with a drift.\\

The diffusive approximation to dipole branching
would be relevant if one were interested in
studying the dipole pair number at a given impact parameter
instead of integrating over all impact parameters, as we do in this paper
since the observable we consider here requires it.
Physically, if the dipole branching process
is diffusive, then the excursions
through very large dipole sizes during evolution
are highly improbable.\\

Let us rewrite the function $\chi(\gamma)$ in the convenient form
$\chi(\gamma)=\lambda-\mu\gamma+\frac{\chi^{\prime\prime}_0}{2}\gamma^2$,
where
\be
\lambda=\chi_0-\gamma_0\chi^\prime_0+\frac12\gamma_0^2\chi_0^{\prime\prime}\ ,
\ \
\mu=\gamma_0\chi_0^{\prime\prime}-\chi_0^\prime.
\ee
With such a kernel, the mean unintegrated 
dipole density is a Gaussian in logarithmic dipole
sizes
\be
f^{(1)}(x;x_{01},y)=\frac{1}{\sqrt{2\pi\chi_0^{\prime\prime}\bar\alpha y}}
\exp\left\{\lambda\bar\alpha y
-\frac{\delta\rho^2}
{2\chi_0^{\prime\prime}\bar\alpha y}
\right\}
\ee
where we introduced 
\be
\delta\rho=\rho-\mu\bar\alpha y\ ,\ \
\rho=\ln \frac{x_{01}^2}{x^2}.
\ee
This formula shows very clearly that viewed from the lines of constant
$\delta\rho$, our process is a branching diffusion with
branching rate $\lambda$ (per unit $\bar\alpha y$) and
diffusion coefficient $\chi^{\prime\prime}_0/2$.
At variance with BFKL, the branching and the diffusion decouple completely
in this class of models.\\

The mean density of pairs of dipoles of
equal size $x$ reads
\be
f^{(2)}(x_{01},x;y)=\langle :\left[f(x_{01},x;y)\right]^2:\rangle
=2\int_0^y \lambda\bar\alpha\, dy_1
\int_{-\infty}^{+\infty}d\rho_1
f^{(1)}(\rho_1;y_1)\left[f^{(1)}(\rho-\rho_1;y-y_1)\right]^2.
\ee
The $\rho_1$ integration is just a Gaussian integration. We are left with
\be
f^{(2)}(x;x_{01},y)=\frac{\lambda}{\pi\chi^{\prime\prime}_0 y}\int_0^y 
\frac{dy_1}{\sqrt{1-y_1^2/y^2}}
\exp\left\{\lambda
\bar\alpha(2y-y_1)
-\frac{\delta\rho^2}
{\chi_0^{\prime\prime}\bar\alpha (y+y_1)}
\right\}.
\ee

We are going to study the large-$y$ limit of $f^{(2)}$.
There are two interesting cases defined by
the relative values of
$|\delta\rho|$ and $\chi_0^{\prime\prime}\bar\alpha y$.\\

Whenever $|\delta\rho|$ is much smaller than $\bar\alpha y$,
the integral over $y_1$ is dominated by a region
near its lower bound.
Indeed, we observe that we can write
\be
f^{(2)}=2\left[f^{(1)}\right]^2\int_0^{\lambda\bar\alpha y}
\frac{d\bar y_1}{\sqrt{1-\bar y_1^2/(\lambda\bar\alpha y)^2}}
\exp\left\{-\bar y_1\left[
1-\frac{1}{\lambda\chi^{\prime\prime}_0}\left(
\frac{\delta\rho}{\bar\alpha y}
\right)^2\frac{1}{1+\bar y_1/(\lambda\bar\alpha y)}
\right]
\right\}
\ee
The argument of the exponential is negative when 
$|\delta\rho|<\sqrt{\lambda\chi^{\prime\prime}_0}\bar\alpha y$.
The fixed-$\delta\rho$ and large-$y$ expansion of $f^{(2)}$
reads
\be
f^{(2)}=2\left[f^{(1)}\right]^2\left[
1+\frac{1}{\lambda\chi_0^{\prime\prime}}
\left(\frac{\delta\rho}{\bar\alpha y}\right)^2
+O\left(1/(\bar\alpha y)^2\right)
\right]
\underset{\bar\alpha y\gg|\delta\rho|}{\longrightarrow }
2\left[f^{(1)}\right]^2
\ee
In this regime, the ``common ancestor'' of the pair of
dipoles lives at the very beginning of the rapidity
evolution, typically within $1/\lambda$ units of $\bar\alpha y$
from the start.
The picture is similar to BFKL either in the DL approximation,
or with a cutoff modeling confinement.\\

In the opposite regime, that is when $\delta\rho\gg\bar\alpha y$,
we write
\be
f^{(2)}=2f^{(1)}\times\lambda
\sqrt{\frac{\bar\alpha y}{2\pi\chi_0^{\prime\prime}}}
\int_0^y
\frac{d{\tilde y_1}}{\sqrt{{\tilde y_1}(2y-{\tilde y_1})}}
\exp\left\{
-{\tilde y_1}\left[
\frac{\delta\rho^2}{4\chi^{\prime\prime}_0\bar\alpha}
\frac{1}{y(y-{\tilde y_1}/2)}
-\lambda\bar\alpha
\right]
\right\},
\label{eq:smally}
\ee
where we defined $\tilde y_1=y-y_1$.
Thus
\be
f^{(2)}\underset{\bar\alpha y\ll|\delta\rho|}{\longrightarrow }
2 f^{(1)}\times\lambda
\frac{\bar\alpha y}{\delta\rho}.
\ee
In this regime, since the low-${\tilde y_1}$ region (i.e.
in terms of the initial integration variable, $y_1\sim y$) 
dominates
the integral in Eq.~(\ref{eq:smally}), the evolution is essentially
a single path until the very last splittings.\\

We see that there is no regime in which a large object can be produced
at the beginning of the evolution, unlike in the plain BFKL case.
This is due to the lack of singularities in the branching-diffusion kernel.
Therefore, the diffusive approximation lacks some fundamental features of the BFKL
evolution, and should be used with great care when one is interested in the
integrated gluon density.\footnote{%
The diffusive approximation is however well-justified for the evolution
of the gluon density at a fixed impact parameter, as was
previously documented (see e.g. Ref.~\cite{Munier:2008cg,Mueller:2010fi}).}


\section{\label{app:num}Numerical simulations in a simplified model}

In order to test qualitatively the analytical results obtained in this paper,
we perform a Monte Carlo simulation of a model that has the main features of the
color dipole model, but that is simpler and more manageable numerically.\\

The model we consider is a simplified version of the color dipole model.
The transverse space has only one dimension, and the evolution kernel
is reduced to the collinear and anticollinear logarithmic
singularities.\\

Let us introduce the logarithmic variable\footnote{
Note that with the full two-dimensional transverse space,
the natural variable is $\rho=\ln x_{01}^2/r^2$.}
$\rho=\ln x_{01}/r$ to characterize a dipole of generic size~$r$.
The equivalent BFKL kernel, that is the rate of splitting 
of a dipole of (log)size $\rho$ 
to a dipole of (log)size $\rho^\prime$, reads
\be
d\rho^\prime\frac{dp}{dy}(\rho\rightarrow\rho^\prime)= d\rho^\prime
\bar\alpha\left[\theta(\rho-\rho^\prime)
+\theta(\rho^\prime-\rho)e^{-(\rho^\prime-\rho)}\right].
\label{eq:collinearkernel}
\ee
The parent dipole remains unchanged.\\

The equivalent BK equation for this model reads
\be
\frac{\partial}{\partial y}Z(\rho,y|u)=\bar\alpha
Z(\rho,y|u)\int_0^{+\infty}d\rho^\prime\, 
\frac{dp}{dy}(\rho\rightarrow\rho^\prime)\left[
Z(\rho^\prime,y|u)-1
\right].
\ee
The eigenfunctions of the kernel are $e^{-\gamma\rho}$, and
the characteristic function is 
\be
\bar\alpha\chi(\gamma)=\bar\alpha
\left(\frac{1}{\gamma}+\frac{1}{1-\gamma}\right).
\ee
One gets the double-log limit by simply turning off the splittings to larger dipoles
(namely by leaving out the second term in the probability 
density $\frac{dp}{dy}(\rho\rightarrow\rho^\prime)$ in Eq.~(\ref{eq:collinearkernel}).)
On the other hand, the infrared cutoff that models confinement 
is implemented by simply removing dipoles created with
a size larger than some given size of the same order as the size
of the initial dipole.\\

We run the Monte Carlo event generator starting from a single dipole
in the three configurations
we study in this paper: full BFKL, BFKL with an infrared cutoff, and DL limit.
We compare the numerical results to the analytical 
formulae~(\ref{eq:distribBFKL}),~(\ref{eq:multi}) and~(\ref{eq:PnDL}).
More precisely, we use the following parametrizations:
\be
\bar n P_n=
N\times
\begin{cases}
\frac{\bar n}{n}e^{-\ln^2 n/(4\bar\alpha y)} & \text{for full BFKL},\\
\frac{2}{c_\text{DL}}
\left(
\frac{n}{c_\text{DL}\bar n}-1
\right)e^{-n/(c_\text{DL}\bar n)} & \text{for the DL limiting model},\\
\frac{1}{c}e^{-n/(c\bar n)} & \text{for BFKL supplemented with a cutoff}.
\end{cases}
\ee

The plot in Fig.~\ref{fig:plotmc} shows the numerical result
for the following parameters: $\bar\alpha y=2.5$ and
$\rho_s=\ln x_{01}/r_s=5$. The fits are performed
for all numerical data points in the range
$n/\bar n\geq 2$. We see that the matching with the
analytical formulae is very good, except, unsurprisingly, for
small values of $n/\bar n$ ($\sim {\cal O}(1)$).
Since full BFKL does not obey KNO scaling, the value of
$\bar n$ is needed in that case: We take it as an output of the
Monte Carlo, $\bar n=3231$.
We get the following determination of the parameters:
\begin{itemize} 
\item BFKL: $N=462$,
\item DL: $N=0.85$,
\item BFKL with a cutoff: $N=1.82$, $c=0.70$.
\end{itemize}
The unnaturally large value of the normalization $N$ in the case
of the BFKL fit is due to the fact that we have neglected
potentially large but slowly varying
factors in the argument of the exponential (see Eq.~(\ref{eq:distribBFKL})), 
of the form $\ln n\times\ln\ln n$.\\

We have not been able to test numerically the universality of $c$ and
the dependence of the prefactor in the BFKL case upon the dipole size.
This would require the use of a Monte Carlo that includes the full QCD dynamics,
such as the one developed in Ref.~\cite{Salam:1996nb},
and to run it in a large enough range of values of $y$ and $\rho$.


\end{document}